\documentclass{ptephy_v1}
\usepackage{bm,latexsym,amssymb,amsmath,mathrsfs}
\usepackage{color}
\usepackage{times,setspace}

\newcommand{\D}{\mathcal{D}}
\renewcommand{\H}{\mathcal{H}}
\renewcommand{\L}{\mathcal{L}}
\renewcommand{\O}{\mathcal{O}}

\begin{document}
\title{Lattice Lindblad simulation}
\author[1]{Tomoya Hayata}
\author[2,3,4]{Yoshimasa Hidaka}
\author[5]{Arata Yamamoto}

\affil[1]{Department of Physics, Keio University, Kanagawa, 223-8521, Japan}
\affil[2]{KEK Theory Center, Tsukuba 305-0801, Japan}
\affil[3]{\it Graduate University for Advanced Studies (Sokendai), Tsukuba 305-0801, Japan}
\affil[4]{\it RIKEN iTHEMS, RIKEN, Wako 351-0198, Japan}
\affil[5]{Department of Physics, The University of Tokyo, Tokyo 113-0033, Japan}

\begin{abstract}
We perform the real-time lattice simulation of an open quantum system, which is based on the Schwinger-Keldysh path integral representation of the Lindblad formalism.
Although the real-time simulation generally suffers from the sign problem, there exist a few exceptional cases.
We focus on a sign-problem-free system of a non-relativistic spinless fermion and analyze time evolution under driving and dissipation.
\end{abstract}

\subjectindex{B38}

\maketitle

\section{Introduction}
Real-time simulation is one of the ultimate goals of lattice gauge theory.
Since the real-time path integral has a complex-valued probability, the Monte Carlo sampling does not work due to the sign problem.
Theoretical physicists have proposed many candidates for the solution, e.g., the complex Langevin method \cite{Berges:2005yt,Berges:2006xc,Fukushima:2014iqa}, the Lefschetz thimble method \cite{Alexandru:2016gsd,Alexandru:2017lqr,Mou:2019tck}, and the tensor network approach \cite{Buyens:2013yza,Pichler:2015yqa,Takeda:2021mnc}.
Algorithms on quantum computers are also being discussed \cite{Martinez:2016yna,Klco:2018kyo,Gustafson:2019vsd,Kharzeev:2020kgc,Yamamoto:2020eqi,Hayata:2021kcp,Gustafson:2021qbt}.
Although these proposals are hopeful, their applications are currently limited to systems with small numbers of degrees of freedom.
Large-scale simulation of non-equilibrium quantum phenomena still remains an open issue.

The Lindblad master equation describes the time evolution of open quantum theory \cite{doi:10.1063/1.522979,Lindblad:1975ef}.
A system gains or loses internal quantities, such as energy-momentum or particle number, through the interaction with an environment. 
In general, the path integral representation of the Lindblad equation is not positive definite.
In very special cases, however, it does not encounter the sign problem.
The known examples are purely dissipative systems \cite{Hebenstreit:2015lwa,Banerjee:2015ela,Caspar:2015ddt,Huffman:2015szi} and a purely fermionic system \cite{Hayata:2021yef,hayata2021dynamical}.
The fermionic system is of particular interest because it can contain both of unitary and dissipative dynamics.
If one can find the system where the fermion determinant is positive definite as a consequence of miracle cancellation, its non-equilibrium property is accessible by the real-time lattice simulation.

In this paper, we study the Lindblad evolution of another fermionic model, which is different from the models in earlier works \cite{Hebenstreit:2015lwa,Banerjee:2015ela,Caspar:2015ddt,Huffman:2015szi,Hayata:2021yef,hayata2021dynamical}.
The lattice formulation is based on the Schwinger-Keldysh path integral representation, which is reviewed in Sec.~\ref{sec2}.
As explicitly shown in Sec.~\ref{sec3}, the model has a significant property; the fermion determinant is exactly quenched and the sign problem is absent.
Owing to this property, the real-time lattice simulation is possible.
We show simulation results in Sec.~\ref{sec4} and discuss theoretical aspects of the model in Sec.~\ref{sec5}.

\section{Path integral formalism of the Lindblad evolution}
\label{sec2}

\begin{figure}[ht]
\begin{center}
 \includegraphics[width=.5\textwidth]{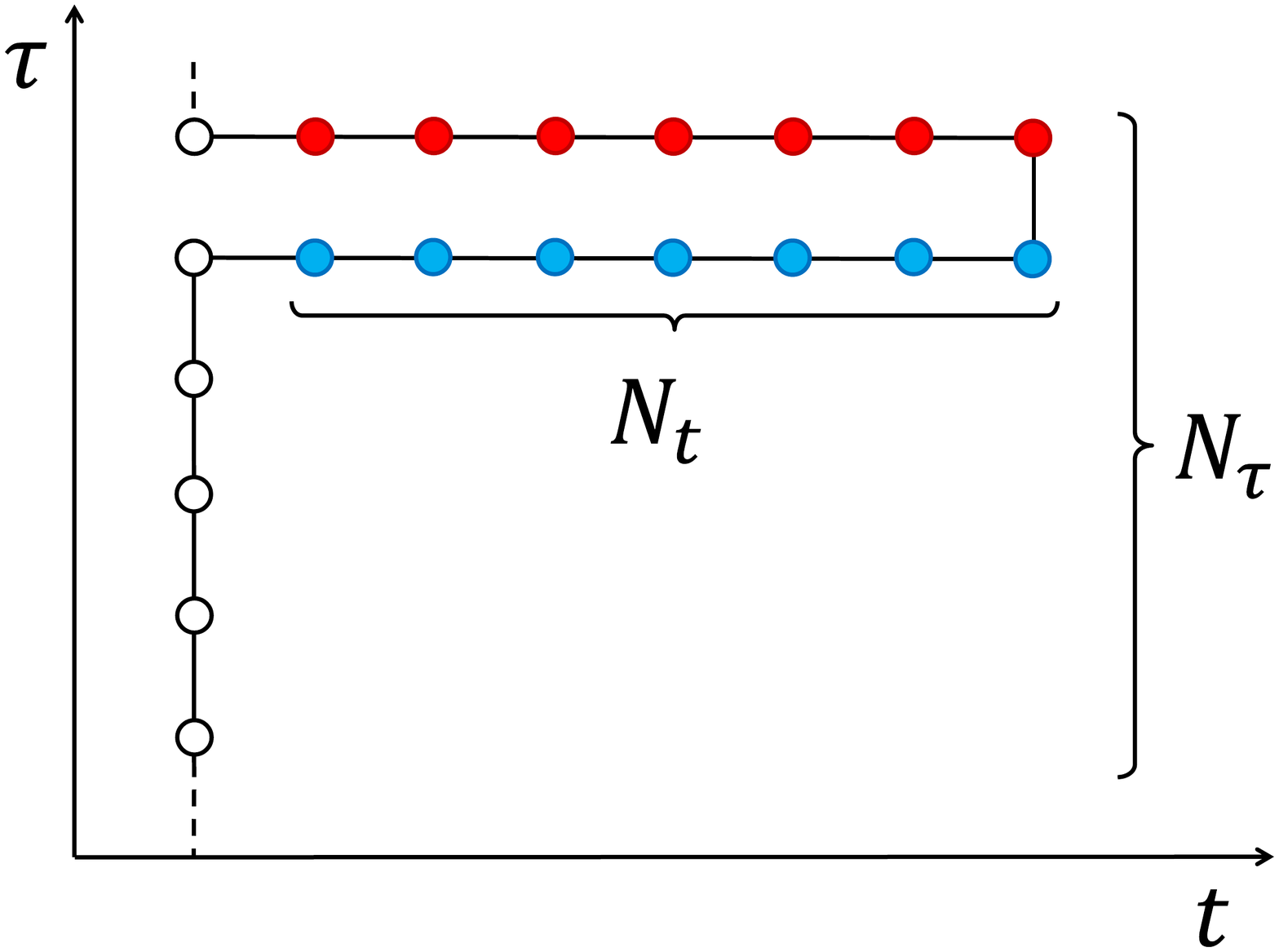}
\caption{
\label{figpath}
Closed path in the real and imaginary time plane.
The red, blue, and white circles stand for $\psi_+$, $\psi_-$, and $\psi_0$, respectively.
The anti-periodic boundary condition is imposed at the ends of the imaginary-time direction.
}
\end{center}
\end{figure}

Let us briefly summarize the path integral representation of the Lindblad formalism.
A comprehensive review can be found in the literature \cite{SBD2016}.
The Lindblad equation was originally formulated in terms of a density matrix in a doubled Hilbert space.
In the path integral representation, it is translated to two branches of the real-time axis: one in the forward direction and the other in the backward direction.
They draw a closed counter, which is called the Schwinger-Keldysh path, in the time plane.
The counter is discretized with the lattice spacing $\delta$.
The fermion field $\psi$ contains three pieces
\begin{equation}
 \psi = 
\begin{cases}
 \psi_+ & (t=\delta,\cdots, N_t\delta)\\
 \psi_- & (t=N_t\delta, \cdots, \delta)\\
 \psi_0 & (t=0)
\end{cases}
\end{equation}
on the discretized time path in Fig.~\ref{figpath}.

The Schwinger-Keldysh path integral is given by
\begin{equation}
 Z= \int \D\psi^* \D\psi \exp\left[iS-S_0\right] .
\end{equation}
For the sake of brevity, we consider a non-relativistic one-component fermion on the three-dimensional spatial lattice.
The continuous-time limit $\delta\to 0$ is assumed to be taken eventually, while the continuous-space limit is not considered here.
The time evolution at $t>0$ is governed by the real-time action
\begin{equation}
\begin{split}
 S=& \int dt \sum_x \bigg[ i \psi_+^* \frac{\partial}{\partial t} \psi_+ - \H_{+} - i \psi_-^* \frac{\partial}{\partial t} \psi_- + \H_{-}
\\
 & - i \sum_k \gamma_k \left\{ \L_{k+}\L_{k-}^* -\frac12 \left( \L_{k+}^*\L_{k+} + \L_{k-}^*\L_{k-} \right)\right\}  \bigg] . 
\end{split}
\end{equation}
The Hamiltonian $\H_\pm$ determines unitary dynamics of $\psi_\pm$.
The second line represents dissipations caused by interactions with an environment.
The index $k$ labels the species of the dissipations, $\gamma_k$ are their strengths, and $\L_{k\pm}$ are the so-called jump operators.
The initial condition at $t=0$ is controlled by the imaginary-time action
\begin{equation}
 S_0= \int d\tau \sum_x \left[ \psi_0^* \frac{\partial}{\partial \tau} \psi_0 + \H_{0} \right]
\end{equation}
with the same Hamiltonian as $\H_\pm$ but of $\psi_0$.
This means that the initial state is a thermal state at the temperature $T=1/(N_\tau\delta)$.

The expectation value of an operator is given by the average of the two branches,
\begin{equation}
 O_c = \frac12 \langle \O_+ + \O_- \rangle,
\label{eqOC}
\end{equation}
where $\O_+$ and $\O_-$ are the corresponding operators on the upper and lower branches.
On the other hand, the difference
\begin{equation}
 O_q = \langle \O_+ - \O_- \rangle
\label{eqOQ}
\end{equation} 
is identically zero for any physical operator \cite{SBD2016}.
The average and difference are called the ``classical'' and ``quantum'' parts of the observable, respectively.

\section{Model without the sign problem}
\label{sec3}

For the unitary part, we consider the tight-binding model on a three-dimensional cubic lattice, 
\begin{align}
 \H_\pm(x) &= \sum_{k} \left[ \{-w+iA_k(x)\} \psi^*_\pm (x) \psi_\pm (x+e_k) + \{-w-iA_k(x)\} \psi^*_\pm (x+e_k) \psi_\pm (x) \right],
\\
 \H_0(x) &= -w \sum_{k} \left[ \psi^*_0 (x) \psi_0 (x+e_k) + \psi^*_0 (x+e_k) \psi_0 (x) \right],
\label{eqh0}
\end{align}
where $w$ is the hopping parameter and $e_k$ is the unit lattice vector in the $x_k$ direction ($k=1,2,3$).
The external field $A_k(x)$ couples to the electric current.
We choose the electric current operators as the jump operators
\begin{equation}
 \L_{k\pm}(x) = j_{k\pm}(x) \equiv i \{ \psi^*_\pm (x) \psi_\pm (x+e_k) - \psi^*_\pm (x+e_k) \psi_\pm (x) \}.
\end{equation}
When the dissipation is isotropic, $\gamma_1=\gamma_2=\gamma_3\equiv \gamma$, the dissipation term is written as the completed square form
\begin{equation}
 - i \sum_k \gamma_k \left\{ \L_{k+}\L_{k-}^* -\frac12 \left( \L_{k+}^*\L_{k+} + \L_{k-}^*\L_{k-} \right)\right\} =  i \sum_k \frac{\gamma}{2} (j_{k+}-j_{k-})^2 .
 \label{eqdiss}
\end{equation}
Here we used the Hermiticity $j_{k\pm}=j_{k\pm}^*$.

After the Hubbard-Stratonovich transformation, the dissipation term is rewritten by the auxiliary field $B_k(x)$ as
\begin{equation}
 Z= \int \D\psi^* \D\psi \D B \exp\left[iS'-S_0-\int dt \sum_x \sum_k \frac{1}{2\gamma} B_{k}^2\right]
\end{equation}
with
\begin{equation}
 S'= \int dt \sum_x \bigg[ i\psi_+^* \frac{\partial}{\partial t} \psi_+ - \H'_{+} - i\psi_-^* \frac{\partial}{\partial t} \psi_- + \H'_{-} \bigg],
\end{equation}
and
\begin{equation}
\begin{split}
 \H'_\pm(x) =& \sum_{k} [ \{-w+iA_k(x)+iB_k(x)\} \psi^*_\pm (x) \psi_\pm (x+e_k) 
\\
&+ \{-w-iA_k(x)-iB_k(x)\} \psi^*_\pm (x+e_k) \psi_\pm (x) ] . 
\label{eqh2}
\end{split}
\end{equation}
Thus, the time evolution is unitary, but it is subject to the Gaussian dephasing noise. 
Since the fermion action is in the bilinear form $\psi^* D \psi$, we can integrate out the fermion field, and then get
\begin{equation}
 Z= \int \D B \ \det D \exp\left[ -\int dt \sum_x \sum_k \frac{1}{2\gamma} B_{k}^2 \right].
\label{eqZB}
\end{equation}
When we choose the backward difference for the time derivative, $\frac{\partial}{\partial t} \psi(t) \equiv \frac{1}{\delta} \{ \psi(t) - \psi(t-\delta) \}$, the matrix representation in the time direction is
\begin{equation}
D=
\begin{pmatrix}
1+i\delta h_1 & 0 & &&&&&&+1\\
\ddots & \ddots & \ddots \\
 & -1 & 1+i\delta h_{N_t} & 0\\
&&-1& 1-i\delta h_{N_t} & 0\\
&&&\ddots&\ddots&\ddots\\
&&&&-1&1-i\delta h_1&0\\
&&&&&-1&1+\delta h_0&0\\
&&&&&&\ddots&\ddots&\ddots\\
&&&&&&&-1&1+\delta h_0
\end{pmatrix}
,
\label{eqD}
\end{equation}
where $h_0$ and $h_n$ $(n=1,\cdots,N_t)$ are the matrix representation of Eqs.~\eqref{eqh0} and \eqref{eqh2}, respectively.
The matrix structure in the spatial directions is omitted in Eq.~\eqref{eqD}.
The fermion determinant is given by
\begin{equation}
 \det D = \det\left[ 1 + \prod_{n=1}^{N_t} (1+i\delta h_n) \prod_{m=N_t}^1 (1-i\delta h_m) ( 1 + \delta h_0 )^{N_\tau} \right].
 \label{eqdet}
\end{equation}
When $N_\tau$ is an even number, this is positive definite because 
\begin{equation}
 ( 1 + \delta h_0 )^{\frac{N_\tau}{2}} \prod_{n=1}^{N_t} (1+i\delta h_n) = \left[ \prod_{m=N_t}^1 (1-i\delta h_m) ( 1 + \delta h_0 )^{\frac{N_\tau}{2}} \right]^\dagger.
\end{equation}
The Monte Carlo sampling is free from the sign problem.

The fermion determinant has a significant property.
We can evaluate the determinant \eqref{eqdet} as
\begin{equation}
 \det D = \det\left[ 1 + (1+O(\delta^2)) ( 1 + \delta h_0 )^{N_\tau} \right] = {\rm const.}+O(\delta^2).
\end{equation}
This is a $B_k$-independent constant in the continuous-time limit $\delta \to 0$.
(Note that the auxiliary field action in Eq.~\eqref{eqZB} is $O(\delta)$.)
This property can be more clearly understood in another discretization manner.
Following the discretization in the time-ordered form \cite{Blankenbecler:1981jt}, we can deform Eq.~\eqref{eqD} as
\begin{equation}
D \simeq
\begin{pmatrix}
1 & 0 & &&&&&&e^{-i\delta h_1}\\
\ddots & \ddots & \ddots \\
 & -e^{-i\delta h_{N_t}} & 1 & 0\\
&&-e^{i\delta h_{N_t}}& 1 & 0\\
&&&\ddots&\ddots&\ddots\\
&&&&-e^{i\delta h_1}& 1 &0\\
&&&&&-e^{-\delta h_0}&1&0\\
&&&&&&\ddots&\ddots&\ddots\\
&&&&&&&-e^{-\delta h_0}&1
\end{pmatrix}
\label{eqD2}
\end{equation}
The two discretizations \eqref{eqD} and \eqref{eqD2} are equivalent up to the higher order of $\delta$, which is irrelevant for the continuous-time limit.
In this form, the determinant is 
\begin{equation}
\det D\simeq \det\left[ 1 + \prod_{n=1}^{N_t} e^{-i\delta h_n} \prod_{m=N_t}^1 e^{i\delta h_m} ( e^{-\delta h_0} )^{N_\tau} \right]
= \det\left[ 1 + e^{-\delta h_0 N_\tau} \right].
\end{equation}
This is completely independent of $B_k$.
We do not need to compute the fermion determinant in the Monte Carlo simulation.
This looks like the famous ``quenched approximation''.
In this model, however, the quenching is not approximate but exact.
This property comes from the specialty of the dissipation term.
Through the Hubbard-Stratonovich transformation, the difference $j_{k+}-j_{k-}$ in the dissipation term is replaced by $B_k$.
As mentioned in Sec.~\ref{sec2}, the quantum part $j_q$ is zero so $\langle B_k \rangle$ must be zero.
If the determinant had a linear term proportional to $B_k$, $\langle B_k \rangle$ would be nonzero.
Therefore, the leading $B_k$-dependent term is not $O(\delta B_k)$ but $O(\delta^2 B_k^2)$ as long as the determinant is a polynomial of the dimensionless combination $\delta B_k$.
This argument holds true for any bilinear Hermitian jump operator written as the completed square form~\eqref{eqdiss}.

\section{Simulation results}
\label{sec4}

We show several results obtained by the simulation.
As mentioned above, the fermion determinant can be quenched and the Monte Carlo weight is the Gaussian distribution.
This property greatly reduces simulation cost.
The simulation is just a white noise sampling and any update algorithm is not necessary.
The hopping parameter is fixed at $w=0.1/\delta$.
All the dimensional quantities in this section are scaled in the unit of $w$.
Spatial lattice volume is $N_s^3=8^3$ and imaginary-time length is $N_\tau=4$.
Since the path integral is the grand canonical ensemble constructed from coherent states, a particle number is defined as an expectation value.
We numerically checked that the expectation value of the particle number density
\begin{equation}
 \bar{n}_c = \frac12 \langle \psi_+^*(t,\vec{x}) \psi_+(t-\delta,\vec{x}) + \psi_-^*(t,\vec{x}) \psi_-(t-\delta,\vec{x}) \rangle
 \label{eqnc}
\end{equation}
is conserved in the real-time evolution.
The time-splitting form of Eq.~\eqref{eqnc} originates from a chemical potential on the lattice \cite{Hasenfratz:1983ba}.

\begin{figure}[ht]
\begin{minipage}[c]{0.5\textwidth}
\begin{center}
 \includegraphics[width=1\textwidth]{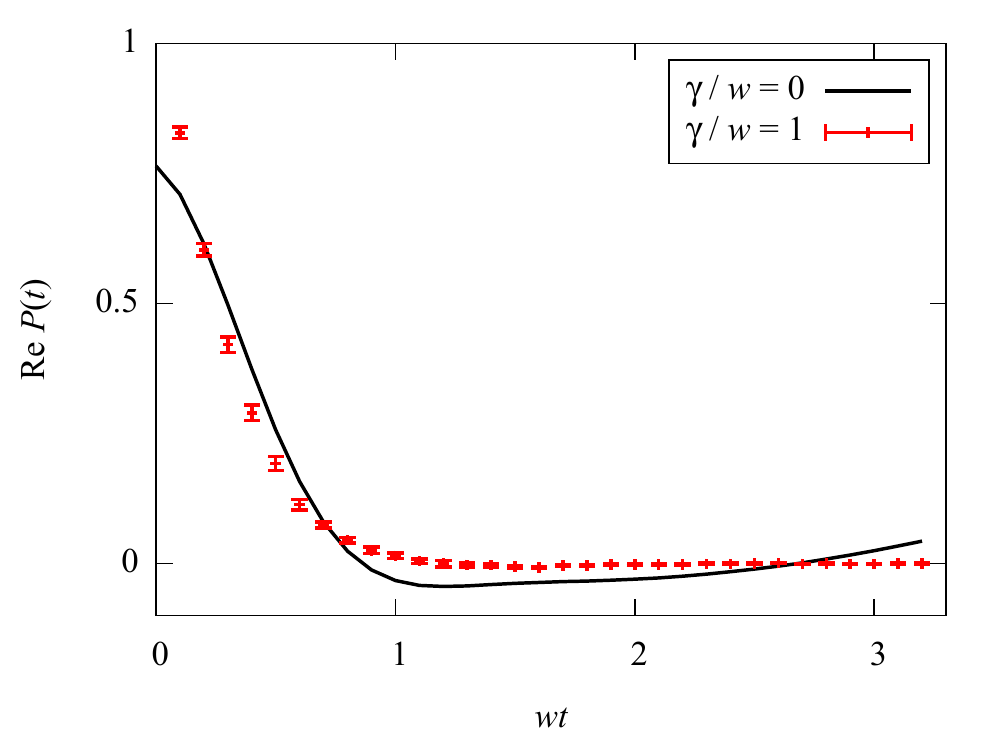}
\end{center}
\end{minipage}
\begin{minipage}[c]{0.5\textwidth}
\begin{center}
 \includegraphics[width=1\textwidth]{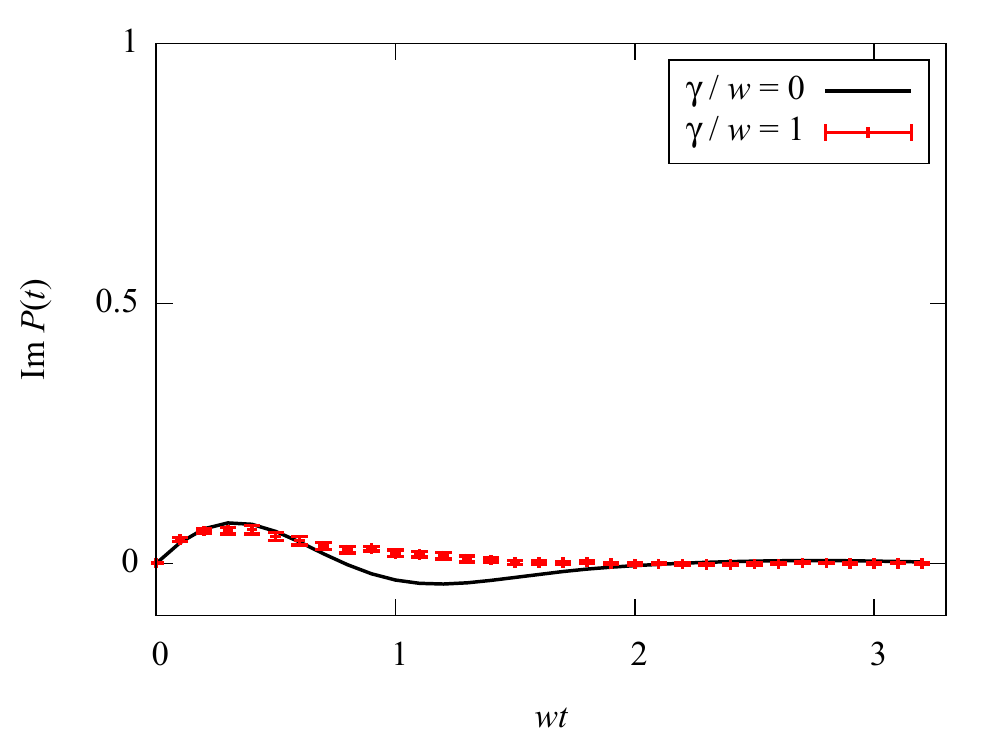}
\end{center}
\end{minipage}
\caption{
\label{figP}
The real-time propagator.
The real-time length is $N_t=32$.
}
\end{figure}

\begin{figure}[ht]
\begin{center}
 \includegraphics[width=0.5\textwidth]{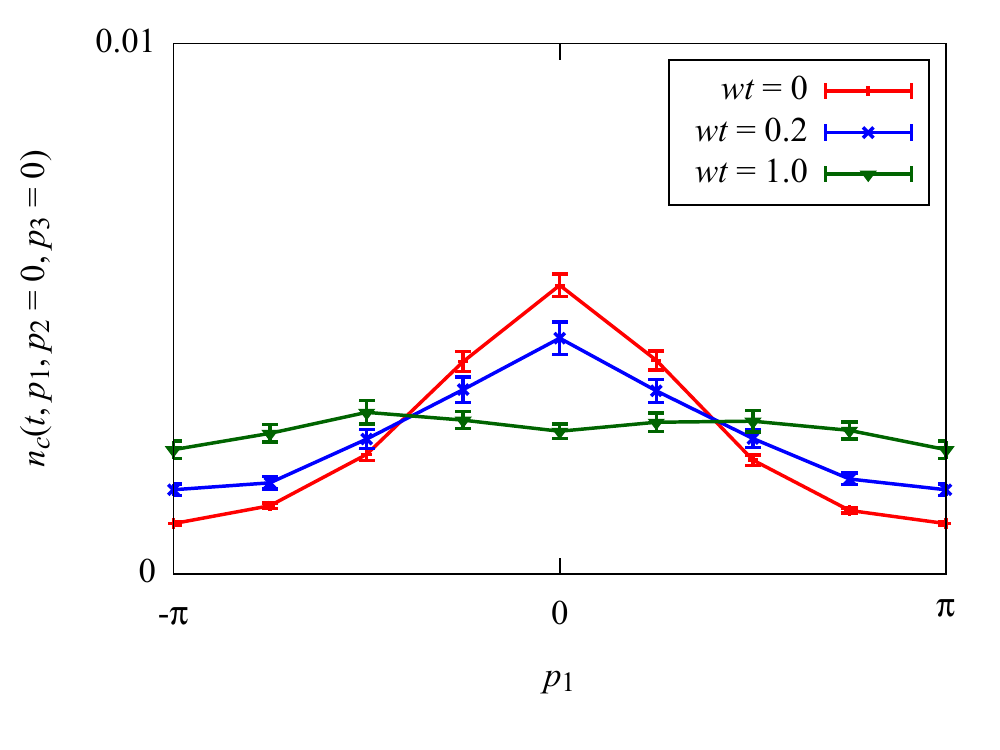}
\end{center}
\caption{
\label{figN}
The real-time evolution of the occupation number.
The data are plotted as functions of the momentum $p_1$ and the other momenta $p_2$ and $p_3$ are set to zero.
The real-time length is $N_t=16$ and the dissipation parameter is $\gamma/w=1$.
}
\end{figure}

A fundamental quantity of interest is the fermion propagator in the real-time direction.
The propagator on the positive branch
\begin{equation}
 P(t) = \left\langle \psi_+(t,\vec{x}) \psi_+^*(0,\vec{x}) \right\rangle
\end{equation}
is plotted in Fig.~\ref{figP}.
The external field is zero, $A_k(x)=0$, here.
Without the dissipation $\gamma=0$, real-time evolution is oscillatory.
Since the initial state is not an eigenstate of the Hamiltonian, the propagator is given by a superposition of many oscillations.
With the dissipation $\gamma \neq 0$, the propagator damps in time.
Quantum coherence is lost under the Gaussian dephasing noise, i.e., quantum fluctuation of the electric current.
The path integral is a functional integral over all possible classical paths.
Even though the total path integral has space-time translation symmetry, each path can break the symmetry.
Individual particles exchange energies and momenta with the environment, and gradually lose their original information.
This picture can be seen from the occupation number
\begin{eqnarray}
 n_c(t,\vec{p}) &=& \frac{1}{N_s^3} \sum_{\vec{x}-\vec{y}}  n_c(t,\vec{x}-\vec{y}) e^{-i\vec{p} \cdot (\vec{x}-\vec{y})} ,
\\
 n_c(t,\vec{x}-\vec{y}) &=& \frac12 \langle \psi_+^*(t,\vec{x}) \psi_+(t-\delta,\vec{y}) + \psi_-^*(t,\vec{x}) \psi_-(t-\delta,\vec{y}) \rangle.
\end{eqnarray}
The momentum integral of the occupation number gives the particle number density, $\bar{n}_c =\sum_{\vec{p}}n_c(t,\vec{p})$, so the integral is conserved while the distribution is not.
As shown in Fig.~\ref{figN}, the occupation number changes from the initial thermal distribution to a uniform one as time goes by.
The original information of the initial state is lost by the dissipation.

\begin{figure}[ht]
\begin{minipage}[c]{0.5\textwidth}
\begin{center}
 \includegraphics[width=1\textwidth]{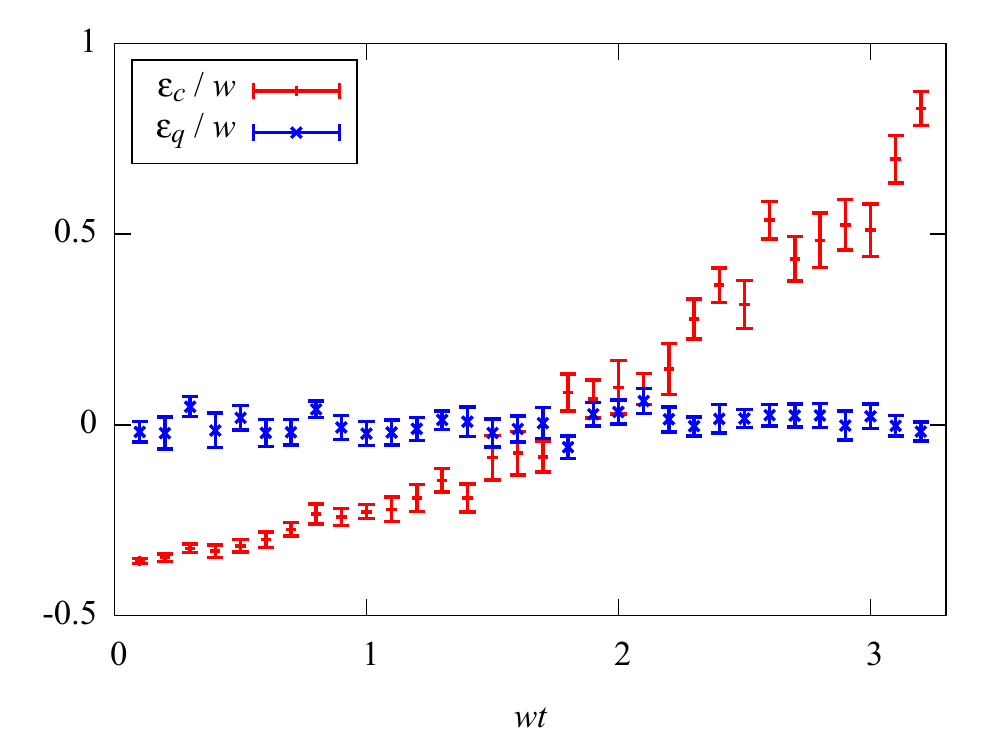}
\end{center}
\end{minipage}
\begin{minipage}[c]{0.5\textwidth}
\begin{center}
 \includegraphics[width=1\textwidth]{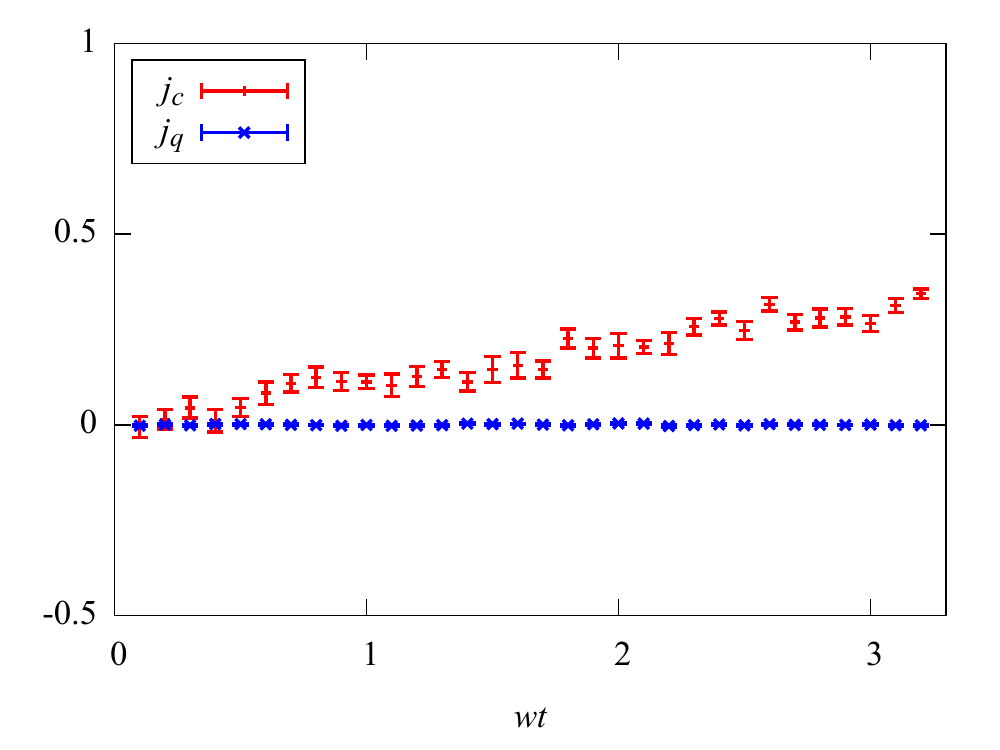}
\end{center}
\end{minipage}
\caption{
\label{figEJ}
The total energy per volume (left) and the electric current (right) under the external electric field.
The real-time length is $N_t=32$ and the dissipation parameter is $\gamma/w=1$.
}
\end{figure}

We can also study the time evolution of physical quantities.
We set the external field
\begin{equation}
 A_1(x) = -Et, \quad A_2(x)=A_3(x)=0
\end{equation} 
such that the constant electric field $E$ drives the electric current in the $x_1$ direction.
The strength of the electric field is set $E/w^2=1$.
The total energy per volume
\begin{equation}
 \varepsilon_c(t) = \frac12 \langle \H_+(x) +\H_-(x) \rangle, \quad \varepsilon_q(t) = -i \langle \H_+(x) - \H_-(x) \rangle,
\end{equation}
and the electric current in the $x_1$ direction
\begin{equation}
 j_c(t) = \frac12 \langle j_{1+}(x) + j_{1-}(x) \rangle, \quad j_q(t) = -i\langle j_{1+}(x) - j_{1-}(x) \rangle
\end{equation}
are plotted in Fig.~\ref{figEJ}.
The factor $-i$ is multiplied to make $\varepsilon_q$ and $j_q$ real.
As the electric field is constantly applied, the electric current $j_c$ is induced and linearly increased.
The system energy $\varepsilon_c$ is also increased.
The energy gain is purely due to the electric field because only the external field explicitly breaks time translation symmetry and other terms, including the dissipation term, do not violate energy conservation.
The quantum parts $\varepsilon_q$ and $j_q$ are always zero within the margin of error as expected.

\section{Discussion}
\label{sec5}

Because of the special property of the fermion determinant, the dissipation term is equivalent to the Gaussian quantum fluctuation.
The resultant theory is very close to the stochastic Schr\"{o}dinger equation.
Particles are stochastically scattered and result in random trajectories.
In this model, the stochastic noise couples to the current, i.e., momentum, so the momentum distribution changes in time while the total momentum is conserved.
It is known that such an environment-induced diffusion leads to an efficient quantum transport~\cite{Rebentrost2009,Plenio2008}. 
Our formulation serves as an efficient numerical method for its large-scale simulation.

The real-time propagator is a damping function.
Since non-local correlation functions are given by the products of the propagator, all the non-local-time correlators will be damping functions, too.
On the other hand, local-time observables of conserved quantities, such as $\varepsilon_c(t)$ and $j_c(t)$, are not damped because the dissipation term does not couple to these observables.
The dissipation couples only to the non-observable $j_q$.
Non-equilibrium steady states cannot be realized by the competition between the dissipation and the driving in this model.
In general, bilinear and Hermitian jump operators can realize only trivial steady states, where the density matrix is unity \cite{Eisler2011,PhysRevA.87.012108}.
We need to discover some other models without the sign problem to study the formation of non-trivial steady states.

As mentioned in Sec.~\ref{sec3}, our method works for any bilinear Hermitian jump operators, although we chose the current operator in this model.
We can generally examine the evolution of an open free-fermion theory under the dephasing process that randomizes the phase of the wave function, and the resultant diffusive dynamics such as the charge transport by real-time lattice simulation.

\ack
This work was supported by JSPS KAKENHI Grant Numbers 19K03841, 21H01007, and 21H01084.
This work was achieved through the use of large-scale computer systems at the Cybermedia Center, Osaka University.

\bibliographystyle{ptephy}
\bibliography{paper}

\end{document}